# Voltage-Independent Active-Power Droop Coefficient for Enhanced Andronov–Hopf Oscillator Grid-Forming Inverters

Hamed Rezazadeh, Mohammad Monfared, *Senior Member, IEEE,*
Meghdad Fazeli, *Senior Member, IEEE*, and Saeed Golestan, *Senior Member, IEEE*

*Abstract*— In recent years, virtual oscillator control, particularly the Andronov-Hopf oscillator (AHO), has received widespread attention for controlling grid-forming (GFM) inverters due to their superior dynamic response. However, traditional AHO systems feature droop coefficients that are dependent on the oscillator voltage amplitude, limiting their ability to maintain consistent grid support during disturbances and resulting in power-sharing inaccuracies. This paper presents an enhanced AHO (EAHO) strategy, where the active power droop coefficient is no longer a function of the voltage amplitude and retains the key dynamic benefits of the original AHO. The EAHO improves both frequency and voltage support and ensures accurate power sharing with other GFM inverters in grid-connected and stand-alone modes. Extensive comparative and small-signal analyses, alongside experimental validation on 2.5 kVA single-phase inverters, confirm the EAHO's improved steady-state performance, enhanced active and reactive power support, and stable operation under varying grid conditions.

## I. INTRODUCTION

WITH the growing dominance of inverter-based resources in modern power systems, grid-forming (GFM) inverters have gained considerable attention for their critical role in maintaining system stability and resilience. A common GFM approach involves synchronous generator (SG) emulation-based methods, which provide grid frequency and voltage support, as well as communication-free power-sharing, by emulating the behaviour of SGs. These methods include droop control [1], virtual synchronous generator [2], [3], synchronverter [4], [5], and synchronous power control [6]. Among these strategies, droop control is widely adopted due to its simplicity, where the angular frequency ($\omega$) and amplitude ($V_p$) of the inverter's voltage are linear functions of active ($P$) and reactive ($Q$) powers, respectively [1].

To improve the dynamic response of the SG-emulation-based methods, virtual oscillator control (VOC) has been proposed as an emerging GFM strategy. VOC leverages the synchronisation behaviour of coupled oscillator networks, allowing the system to quickly stabilise to a limit cycle [7], [8]. As a result, it achieves global asymptotic synchronisation and enhanced transient performance compared to traditional droop control [9]. Furthermore, VOC's control strategy relies directly on the inverter output current, eliminating the need for power measurements and low-pass filtering, thereby further improving transient speed and power-sharing dynamics [10]. Among various VOC implementations, the Andronov-Hopf oscillator (AHO) stands out for its dispatchability and ability to generate a harmonic-free voltage waveform [11], [12], [13].

Recently, the AHO has been equipped with virtual inertia, increasing its relevance in modern power systems, although at the cost of reduced damping [14], [15]. To address this issue, two recent damping enhancement methods have been proposed [16], [17]. Additionally, a new control strategy in the dq reference frame has been developed to enhance the power tracking and dynamic response of the VOC [18]. The transient stability of the AHO has also been improved through modifications to its control structure [19] or by optimising its control parameters [20].

Despite these advantages, the droop coefficients of the AHO are nonlinear functions of the voltage amplitude $V_p$. In particular, the active power loop (APL) droop coefficient, unlike in conventional droop control, varies with $V_p$, complicating the design of control parameters, as they must account for the allowable range of $V_p$. As a result, the inverter's ability to provide consistent grid support during frequency disturbances can be compromised. Moreover, power-sharing inaccuracies may arise when AHO-based inverters operate in parallel with other GFM units. To address these challenges, a modified approach has been proposed in [21] to achieve linear droop functionality by adjusting the AHO's core structure. However, as stated in [19], adjusting the core structure of the AHO compromises its dynamic performance.

This paper proposes an enhanced AHO (EAHO) control strategy that achieves active power droop functionality

University, SA1 8EN Swansea, U.K. (e-mail: 2251309@swansea.ac.uk; mohammad.monfared@swansea.ac.uk; m.fazeli@swansea.ac.uk).
Saeed Golestan is with AAU Energy, Aalborg University, 9220 Aalborg East, Denmark (e-mail: sgd@energy.aau.dk).

equivalent to conventional droop control, while preserving the AHO's core structure and its fast dynamics. The proposed strategy improves grid support capabilities in response to fluctuations in both grid frequency and voltage amplitude and ensures accurate power-sharing in stand-alone mode, even when operating in parallel with other GFM strategies. Here, the EAHO is designed for single-phase applications to address a gap arising from the increasing deployment of small-scale distributed generation in residential and commercial areas, mainly rooftop solar PVs and energy storage systems, which necessitates the use of single-phase GFM inverters to ensure safe and stable operation of inverter-based resources [22], [23], [24]. Nevertheless, the strategy is scalable and applicable to three-phase systems. Furthermore, although the focus is on $\omega$-$P$ and $V_p$-$Q$ droop functionality, the proposed strategy is also compatible with dominantly resistive systems with $\omega$-$Q$ and $V_p$-$P$ droop. The effectiveness of the proposed strategy is demonstrated through large-signal and small-signal analysis with experimental validation on 2.5 kVA single-phase inverters in both grid-connected and stand-alone modes.

The rest of this paper is organised as follows: Section II reviews the conventional AHO and introduces the proposed EAHO strategy, accompanied by a comprehensive analysis and a small-signal state-space stability investigation. Section III presents the experimental validation, and Section IV concludes the paper.

## II. CONVENTIONAL AND ENHANCED AHO

### A. Conventional AHO

The AHO is a time-domain controller implemented in the $\alpha\beta$ reference frame. The control law for the oscillator output voltage is expressed as:

$$\dot{v} = \underbrace{j\omega_0 v}_{1} + \underbrace{\mu\left(V_{p0}^2 - V_p^2\right)v}_{2} + \underbrace{j\,\eta(i_{ref} - i)}_{3} \qquad (1)$$

where $v$ is the oscillator output voltage, $\omega_0$ and $V_{p0}$ represent the nominal values for angular frequency ($\omega$) and voltage vector magnitude ($V_p$), $i_{ref}$ is the reference value of the inverter's output current ($i$), and $\eta$ and $\mu$ are AHO's control parameters [12]. This equation illustrates that the AHO dynamics consist of three terms: (1) the harmonic oscillator, (2) the voltage magnitude correction, and (3) the synchronisation feedback (current input) [12]. Equation (1) can be rewritten in the $\alpha\beta$ reference frame as:

$$\begin{bmatrix} \dot{v}_\alpha \\ \dot{v}_\beta \end{bmatrix} = \begin{bmatrix} \mu\left(V_{p0}^2 - V_p^2\right) & -\omega_0 \\ \omega_0 & \mu\left(V_{p0}^2 - V_p^2\right) \end{bmatrix} \begin{bmatrix} v_\alpha \\ v_\beta \end{bmatrix} + \eta \begin{bmatrix} 0 & -1 \\ 1 & 0 \end{bmatrix} \begin{bmatrix} i_{\alpha ref} - i_\alpha \\ i_{\beta ref} - i_\beta \end{bmatrix} \qquad (2)$$

where $v_\alpha$ and $v_\beta$ are the oscillator output voltages, $i_{\alpha ref}$ and $i_{\beta ref}$ are the inverter's reference currents, and $i_\alpha$ and $i_\beta$ denote the inverter's output currents in the $\alpha\beta$ reference frame. In single-phase systems, since only the $\alpha$-axis current is physically available, the current component $i_\beta$ is usually generated by a second-order generalised integrator-based quadrature signal generator (SOGI-QSG) [25].

By defining the oscillator's voltage amplitude and phase as $V_p = \sqrt{v_\alpha^2 + v_\beta^2}$ and $\theta = \arctan(v_\beta/v_\alpha)$, their time derivatives are given by:

$$\begin{cases} \dot{V}_p = \dfrac{v_\alpha \dot{v}_\alpha + v_\beta \dot{v}_\beta}{V_p} \\ \dot{\theta} = \dfrac{\dot{v}_\beta v_\alpha - \dot{v}_\alpha v_\beta}{V_p^2}. \end{cases} \qquad (3)$$

Substituting (2) into (3) yields:

$$\begin{cases} \dot{V}_p = \mu\left(V_{p0}^2 - V_p^2\right)V_p + \dfrac{2\eta}{V_p}(Q_{ref} - Q) \\ \dot{\theta} = \omega = \omega_0 + \dfrac{2\eta}{V_p^2}\left(P_{ref} - P\right) \end{cases} \qquad (4)$$

where $P_{ref}$ and $Q_{ref}$ are the reference values of the active and reactive powers, respectively. As seen from (4), the oscillator's voltage amplitude evolves dynamically until equilibrium is reached ($\dot{V}_p$=0). At this steady state, the droop characteristics of AHO can be obtained as shown in (5).

$$\begin{cases} V_p^4 = V_{p0}^2 V_p^2 + \dfrac{2\eta}{\mu}\left(Q_{ref} - Q\right) \\ \omega = \omega_0 + \dfrac{2\eta}{V_p^2}\left(P_{ref} - P\right) \end{cases} \qquad (5)$$

The APL and reactive power loop (RPL) droop coefficients, $m_p$ and $m_q$, are defined as [9]:

$$\begin{cases} m_p = -\dfrac{d\omega}{dP} \\ m_q = -\dfrac{dV_p}{dQ} = -\dfrac{1}{\dfrac{dQ}{dV_p}}. \end{cases} \qquad (6)$$

Substituting (5) into (6) yields:

$$\begin{cases} m_{p,AHO} = \dfrac{2\eta}{V_p^2} \\ m_{q,AHO} = \dfrac{\eta}{\mu V_p(2V_p^2 - V_{p0}^2)} \end{cases} \qquad (7)$$

where $m_{p,AHO}$ and $m_{q,AHO}$ are the AHO's droop coefficients. From (7), it is evident that $m_{p,AHO}$ is inversely proportional to the square of the voltage amplitude $V_p$. Furthermore, the expression for $m_{q,AHO}$ becomes negative when $V_p < V_{p0}/\sqrt{2}$, which can jeopardise system stability.

According to the droop control principle, active and reactive powers should reach their nominal values ($P_0$ and $Q_0$) in response to the maximum allowable grid frequency deviation ($\Delta\omega_{max}$) and maximum voltage amplitude ($V_{p,max}$), respectively. Substituting these conditions into (5) gives the control parameters as [12]:

$$\eta = \frac{\Delta\omega_{max} V_{p,max}^2}{2P_0}, \quad \mu = \frac{2\eta Q_0}{V_{p,max}^4 - V_{p0}^2 V_{p,max}^2}. \qquad (8)$$

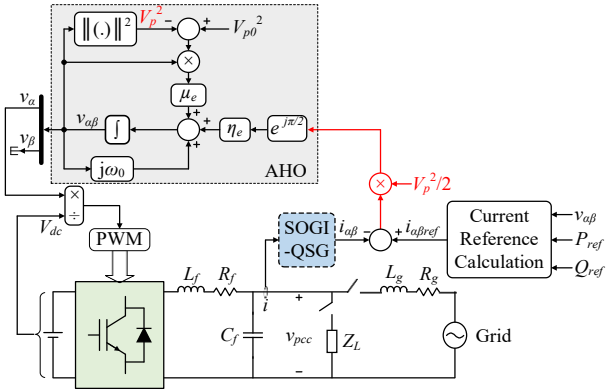

Fig. 1. Proposed EAHO-based single-phase GFM inverter.

## B. Enhanced AHO

Fig. 1 illustrates the proposed EAHO-based GFM scheme for a single-phase inverter, where $L_f$, $R_f$, and $C_f$ denote the filter inductance, its parasitic resistance, and the filter capacitance, respectively. $Z_L$ is the local load, and $L_g$ and $R_g$ represent the grid impedance. In this configuration, $V_{dc}$ is the DC-link voltage and $v_{pcc}$ represents the voltage at the point of common coupling.

As already stated in [19], the first two terms of (1) should remain unchanged to preserve the fast dynamic response and low harmonic distortion of the AHO. Thus, in this work, only the third term in (1) is manipulated as the current error is multiplied by $V_p^2/2$, as highlighted in Fig. 1. The control law of the EAHO can then be expressed as:

$$\begin{bmatrix} \dot{v}_\alpha \\ \dot{v}_\beta \end{bmatrix} = \begin{bmatrix} \mu_e \left( V_{p0}^2 - V_p^2 \right) & -\omega_0 \\ \omega_0 & \mu_e \left( V_{p0}^2 - V_p^2 \right) \end{bmatrix} \begin{bmatrix} v_\alpha \\ v_\beta \end{bmatrix} + \eta_e \frac{V_p^2}{2} \begin{bmatrix} 0 & -1 \\ 1 & 0 \end{bmatrix} \begin{bmatrix} i_{\alpha ref} - i_\alpha \\ i_{\beta ref} - i_\beta \end{bmatrix} \quad (9)$$

where $\eta_e$ and $\mu_e$ are the EAHO's control parameters. Comparing (2) and (9) shows that the first terms are identical, while the current error term has been modified to achieve the APL droop coefficient independent of the voltage amplitude. Substituting (9) into (3) yields:

$$\begin{cases} \dot{V}_p = \mu_e \left( V_{p0}^2 - V_p^2 \right) V_p + \eta_e V_p (Q_{ref} - Q) \\ \dot{\theta} = \omega = \omega_0 + \eta_e (P_{ref} - P). \end{cases} \quad (10)$$

At equilibrium ($\dot{V}_p$=0), the droop characteristics of EAHO can be obtained as (11).

$$\begin{cases} V_p^2 = V_{p0}^2 + \dfrac{\eta_e}{\mu_e} (Q_{ref} - Q) \\ \omega = \omega_0 + \eta_e (P_{ref} - P) \end{cases} \quad (11)$$

As shown in (11), the proposed modification leads to a linear APL droop equation that is independent of $V_p$. Additionally, the RPL droop reduces to a quadratic voltage dependence, rather than the quartic dependence observed in the AHO. Similar to the AHO control parameters, $\eta_e$ and $\mu_e$ should be designed such that the active and reactive powers reach their nominal values at the maximum allowable grid frequency deviation and maximum voltage amplitude, respectively. These parameters can thus be determined from (11), as given in (12).

## TABLE I
### EXPERIMENT PARAMETERS

| Parameters | Description | Value |
|---|---|---|
| $P_0, Q_0$ | rated active and reactive power | 2000 W, 1500 var |
| $V_{dc}$ | DC-link voltage | 380 V |
| $\omega_0$ | nominal angular frequency | $2\pi \times 50$ rad/s |
| $V_{p0}$ | nominal voltage amplitude | 311 V |
| $L_f, C_f$ | filter impedance | 7 mH, 3.9 μF |
| $L_g, R_g$ | grid impedance | 1 mH, 1 Ω |
| $\eta, \mu$ | AHO control parameters | 91.99, $1.16 \times 10^{-4}$ |
| $\eta_e, \mu_e$ | EAHO control parameters | 0.0016, $1.16 \times 10^{-4}$ |
| $m_{p,droop}, m_{q,droop}$ | droop control parameters | 0.0016, 0.0207 |
| $\omega_p, \omega_q$ | LPF bandwidth | 20, 20 rad/s |

$$\eta_e = \frac{\Delta \omega_{max}}{P_0}, \quad \mu_e = \frac{\eta_e Q_0}{V_{p,max}^2 - V_{p0}^2} \quad (12)$$

These parameters are designed based on grid requirements. According to the National Grid Electricity System Operator for Great Britain, the maximum allowable frequency deviation is ±0.5 Hz (i.e., $\Delta f_{max}$=0.01 pu) [26]. Additionally, the typical voltage tolerance is up to 110% of the nominal value. Therefore, the control parameters for both the conventional and proposed strategies are selected using equations (8) and (12), and are summarised in Table I to satisfy these requirements.

The EAHO's droop coefficients can be calculated by substituting (11) into (6), resulting in:

$$\begin{cases} m_{p,EAHO} = \eta_e \\ m_{q,EAHO} = \dfrac{\eta_e}{2\mu_e V_p} \end{cases} \quad (13)$$

where $m_{p,EAHO}$ and $m_{q,EAHO}$ are the droop coefficients under the EAHO strategy. Equation (13) shows that the APL droop equation simplifies to $\eta_e$, making it independent of voltage magnitude. Additionally, in contrast to the AHO case, $m_{q,EAHO}$ remains positive for all $V_p$, thus improving RPL's stability under varying grid conditions.

For the conventional droop control, the control law is given in (14) [1],

$$\begin{cases} V_p = V_{p0} + m_{q,droop}(Q_{ref} - Q_f) \\ \omega = \omega_0 + m_{p,droop}(P_{ref} - P_f) \end{cases} \quad (14)$$

where $P_f$ and $Q_f$ can be obtained using a low-pass filter (LPF) for the active and reactive power to mitigate the double-line frequency oscillation and measurement noises as:

$$\begin{cases} P_f(s) = \dfrac{\omega_p}{s + \omega_p} P \\ Q_f(s) = \dfrac{\omega_q}{s + \omega_q} Q \end{cases} \quad (15)$$

where $\omega_p$ and $\omega_q$ are the LPF cutoff frequencies. Fig. 2 compares the APL droop coefficients for the AHO, EAHO, and conventional droop controllers as functions of $V_p$, using the parameters listed in Table I. The results show that both the EAHO and conventional droop achieve a voltage-independent droop gain, while the AHO's droop coefficient varies considerably, increasing by approximately 21% at nominal voltage. This variation leads to degraded frequency support and inaccurate power sharing during parallel operation under the nominal voltage. Given the limited overcurrent tolerance of power electronic components, such mismatches may result in inverter overloading or even damage.

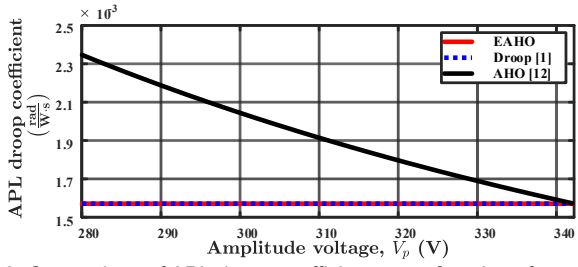

Fig. 2. Comparison of APL droop coefficients as a function of $V_p$.

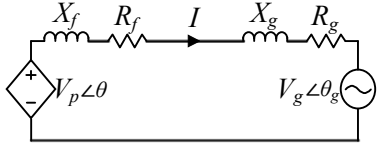

Fig. 3. Simplified model of a single-phase inverter connected to the grid.

In addition to enhancing the APL droop, the proposed strategy also improves the reactive power support capability compared to the conventional AHO. Fig. 3 presents a simplified model of a single-phase inverter connected to the grid. Because of its small value, the effect of the filter capacitance is neglected. Furthermore, within the frequency range of interest, delays introduced by digital control and pulse-width modulation are also considered negligible [12]. Consequently, the controller's output voltage can be approximated as the inverter's terminal voltage.

By defining $\delta=\theta-\theta_g$ as the phase angle difference between the oscillator voltage and grid voltage, the dynamic equations of EAHO, as given in (10), simplify to:

$$\begin{cases} \dot{V}_p = \mu_e\left(V_{p0}^2 - V_p^2\right)V_p + \eta_e V_p(Q_{ref} - Q) \\ \dot{\delta} = \eta_e(P_{ref} - P) \end{cases} \quad (16)$$

According to Fig. 3, the injected active and reactive powers from the inverter to the grid can be calculated as:

$$\begin{cases} P = \dfrac{R_T\left(V_p^2 - V_p V_g\, cos(\delta)\right) + X_T V_p V_g\, sin(\delta)}{2\left(R_T^2 + X_T^2\right)} \\ Q = \dfrac{X_T\left(V_p^2 - V_p V_g\, cos(\delta)\right) - R_T V_p V_g\, sin(\delta)}{2\left(R_T^2 + X_T^2\right)} \end{cases} \quad (17)$$

where $R_T=R_f+R_g$ and $X_T=X_f+X_g$. These expressions confirm that both power components are inherently dependent on voltage magnitude and phase difference, as well as circuit parameters. However, the primary contribution of this paper is that the active power droop coefficient is not any more a function of the voltage amplitude and therefore not affected by its variations.

Substituting (17) into (16) yields a set of two nonlinear differential equations. These equations are solved numerically (using MATLAB) for the parameters given in Table I, assuming a 20% grid voltage drop, with $P_{ref} = 0$ W, and $Q_{ref} = 0$. Under these conditions, the EAHO injects 1443 var of reactive power at steady-state. A similar large-signal analysis for the conventional AHO and droop controllers results in 1078 var and 1529 var, respectively. These results indicate that the proposed EAHO injects approximately 25% more reactive power than the conventional AHO, thereby offering enhanced grid support under voltage amplitude variations. The droop controller delivers slightly more reactive power than the EAHO in this case.

### C. Stability Analysis

This section investigates the stability performance of different control strategies. To this end, the small-signal models of all controllers are derived in state-space form:

$$\Delta \dot{x} = A\,\Delta x + B\,\Delta u \quad (18)$$

where $x$ is the state vector, $u$ represents the system input, $\Delta x$ is a small perturbation around the equilibrium point $X_{eq}$, and $A$ and $B$ are the Jacobian Matrices. The system is exponentially stable if all eigenvalues of matrix $A$ lie in the left half of the complex plane. In the following, $x_d$ and $x_q$ denote the component of $x$ projected onto the grid voltage's $dq$ frame.

Since the grid voltage's $dq$ frame is chosen as the reference, the inverter's voltage components are defined as $v_d=V\cos(\theta)$ and $v_q=V\sin(\theta)$, where $V=V_p/\sqrt{2}$ is the RMS value of the voltage. Thus, for the system model shown in Fig. 3, the current dynamics and output power expressions in the $dq$ frame are given by:

$$\begin{cases} \dot{i}_d = -\dfrac{R_T}{L_T}i_d + \omega i_q + \dfrac{V\cos(\theta)}{L_T} - \dfrac{V_g}{L_T} \\ \dot{i}_q = -\omega i_d - \dfrac{R_T}{L_T}i_q + \dfrac{V\sin(\theta)}{L_T} \end{cases} \quad (19)$$

$$\begin{cases} P = V\cos(\theta)\, i_d + V\sin(\theta)\, i_q \\ Q = V\sin(\theta)\, i_d - V\cos(\theta)\, i_q \end{cases} \quad (20)$$

The dynamic equations of the EAHO were previously derived in (10). By substituting (20) into (10) and linearising the resulting equations and (19) around the equilibrium point, a state-space representation as in (18) is obtained, where $\Delta x=[\Delta v, \Delta\theta, \Delta i_d, \Delta i_q]^{\mathsf{T}}$ and $\Delta u=\Delta v_g$. The resulting Jacobian matrix A is presented in (21), where the equilibrium point is given by $X_{eq}=[V_{eq}, \theta_{eq}, I_d, I_q]^{\mathsf{T}}$ and $\omega_{eq}=\omega_g$. To determine $X_{eq}$, the steady-state conditions $\dot{i}_d=\dot{i}_q=\dot{V}=0$ are applied. The resulting nonlinear algebraic equations are solved using Newton's iteration

$$A = \begin{bmatrix} 2\mu_e\left(V_0^2 - 3V_{eq}^2\right) + \eta_e Q_{ref} - 2\eta_e V_{eq}F_2^* & -\eta_e V_{eq}^2 F_1^* & -\eta_e V_{eq}^2 \sin(\theta_{eq}) & \eta_e V_{eq}^2 \cos(\theta_{eq}) \\ -\eta_e F_1^* & \eta_e V_{eq}F_2^* & -\eta_e V_{eq}\cos(\theta_{eq}) & -\eta_e V_{eq}\sin(\theta_{eq}) \\ \dfrac{\cos(\theta_{eq})}{L_T} & -\dfrac{V_{eq}\sin(\theta_{eq})}{L_T} & -\dfrac{R_T}{L_T} & \omega_{eq} \\ \dfrac{\sin(\theta_{eq})}{L_T} & \dfrac{V_{eq}\cos(\theta_{eq})}{L_T} & -\omega_{eq} & -\dfrac{R_T}{L_T} \end{bmatrix} \quad (21)$$

$$F_1^* = I_d\cos(\theta_{eq}) + I_q\sin(\theta_{eq}), \quad F_2^* = I_d\sin(\theta_{eq}) - I_q\cos(\theta_{eq})$$

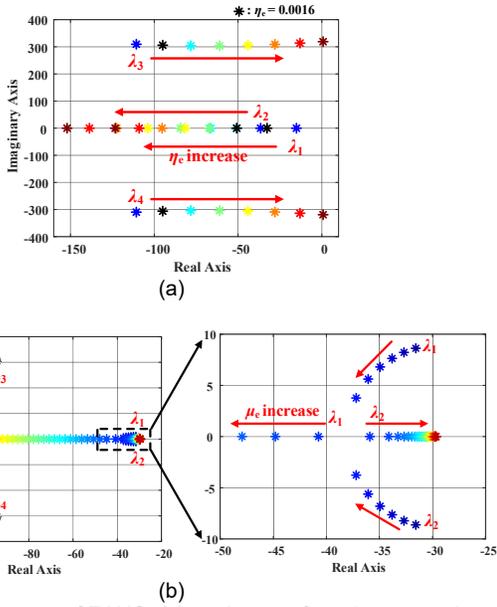

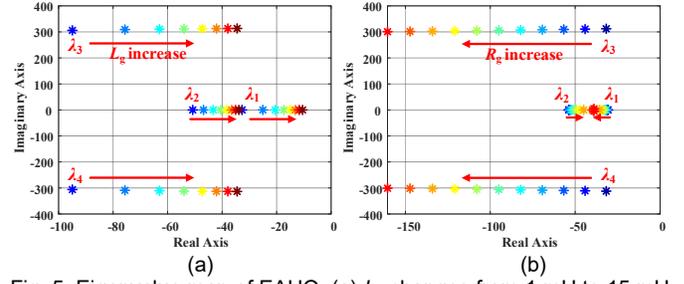

Fig. 5. Eigenvalue map of EAHO: (a) $L_g$ changes from 1 mH to 15 mH and (b) $R_g$ changes from 0.5 Ω to 1.5 Ω.

Fig. 4. Eigenvalue map of EAHO: (a) $\eta_e$ changes from $0.5\eta_{e,set}$ to $4\mu_{e,set}$ and (b) $\mu_e$ changes from $0.1\mu_{e,set}$ to $4\mu_{e,set}$.

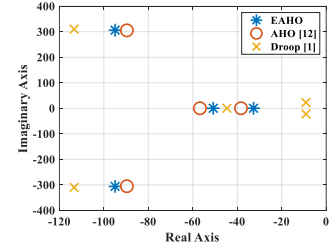

Fig. 6. Eigenvalue comparison.

method, implemented in MATLAB. For the experimental parameters listed in Table I, the equilibrium point is found to be: $X_{eq} = [224.39 \text{ V}, 0.1079 \text{ rad}, 8.72 \text{ A}, 2.24 \text{ A}]^T$.

Fig. 4(a) shows the eigenvalue map of the EAHO as $\eta_e$ changes from $0.5\eta_{e,set}$ to $4\mu_{e,set}$, where $\eta_{e,set} = 0.0016$ is the designed value to satisfy the grid's requirements. This figure reveals that for small values of $\eta_e$, the dominant eigenvalues are $\lambda_1$ and $\lambda_2$, which move leftward, away from the imaginary axis, as $\eta_e$ increases, indicating an improved dynamic response. However, for excessively large $\eta_e$, eigenvalues $\lambda_3$ and $\lambda_4$ become dominant and approach the imaginary axis, suggesting reduced stability margins. For this system, $\eta_e = 4\eta_{e,set}$ marks the critical threshold beyond which the system becomes unstable. The eigenvalues corresponding to $\eta_e = \eta_{e,set}$ (the value dictated by grid requirements), are marked with the black "✳", confirming the stable operation of EAHO.

Fig. 4(b) presents the eigenvalue map as $\mu_e$ changes from $0.1\mu_{e,set}$ to $4\mu_{e,set}$, where $\mu_{e,set}$ is the designed value. This figure reveals that the eigenvalues $\lambda_3$ and $\lambda_4$ change slightly with the change in $\mu_e$. In contrast, $\lambda_1$ moves to the left and away from the imaginary axis, which implies a faster response. However, $\lambda_2$ first moves to the left for $0.1\mu_{e,set} < \mu_e < 0.6\mu_{e,set}$, then changes slightly with the increase of $\mu_e$.

Fig. 5 analyses the performance of the proposed strategy under varying grid impedance. Fig. 5(a) presents the eigenvalue map when $L_g$ changes from 1 mH to 15 mH. This figure indicates that, as $L_g$ increases (weaker grids), all eigenvalues shift closer to the imaginary axis, implying a slower response. Fig. 5(b) illustrates the eigenvalue map when $R_g$ varies from 0.5 Ω to 1.5 Ω (lower $X/R$ ratios). This figure reveals that the dominant eigenvalues $\lambda_1$ and $\lambda_2$ change slightly with $R_g$. However, $\lambda_3$ and $\lambda_4$ shift to the left and move away from the imaginary axis as $R_g$ increases, indicating a faster response. However, for larger values of $R_g$, these eigenvalues become non-dominant, meaning that further increases in $R_g$ have minimal impact on the transient behaviour.

Using a similar approach, the state-space representations and corresponding matrix **A** for the AHO and droop controllers are derived, and reported in Appendix A. Fig. 6 compares the eigenvalue plots of the EAHO, AHO, and droop control under the experimental parameters. The eigenvalues of the EAHO and AHO are close, confirming their similar transient performance. In contrast, the droop control exhibits dominant eigenvalues closer to the imaginary axis, which implies a slower dynamic response.

### D. Step-by-step parameter design

In the following, a straightforward design algorithm is provided, considering both droop functionality and stability margins.

1) **Define system requirements:**
- Rated power of the inverter ($P_0$ and $Q_0$)
- Maximum allowable frequency deviation ($\Delta\omega_{max}$)
- Maximum allowable voltage amplitude ($V_{p,max}$)

2) **Calculate initial control parameters to satisfy droop requirements:**
Use (12) to compute $\eta_e$ and $\mu_e$. For the experimental parameters, these values are $\eta_e = 0.0016$ and $\mu_e = 1.16 \times 10^{-4}$.

3) **Check the small-signal stability for the calculated $\eta_e$ and $\mu_e$:**
From the eigenvalue map in Fig. 4, for the system to remain stable:
- $\eta_e < 0.0062$, which is satisfied (if not, $\eta_e$ must be curtailed to 0.0062).
- $0 < \mu_e$, which is always satisfied.

## III. EXPERIMENTAL VALIDATION

To validate the theoretical findings, a 2.5 kVA test setup is implemented, as shown in Fig. 7, with parameters listed in Table I. The Cinergia 7.5 kVA grid emulator is employed to simulate the grid behaviour. The control algorithms are implemented on a dSPACE DS1007 processor. The effectiveness of the proposed strategy compared to the conventional methods is validated through six test scenarios.

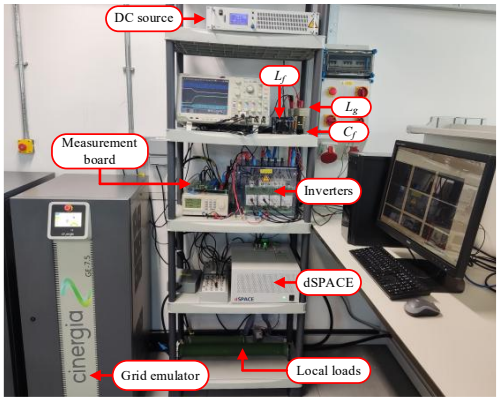

Fig. 7. Experimental test setup.

## A. Test Scenario 1: Grid frequency change

This scenario compares the grid support capabilities in response to grid frequency variations. Here, the grid frequency decreases from 50 Hz to 49.5 Hz, with $P_{ref} = 0$ W. According to the droop performance and parameter design, the inverters are expected to inject nominal active power (2000 W). Fig. 8 compares the experimental results for the proposed EAHO and conventional AHO and droop controllers. As shown in Figs. 8(a) and (c), both the EAHO and droop controllers successfully inject 2000 W, whereas Fig. 8(b) shows the AHO injecting only 1800 W, 10% below the target value. This lower active power injection by the AHO is consistent with the behaviour predicted in Fig. 2, where its APL droop coefficient increases at nominal voltage.

## B. Test Scenario 2: Grid voltage amplitude change

This scenario compares the inverters' reactive power support when the grid voltage first falls by 20% and then jumps by 10%, with $P_{ref}$ and $Q_{ref}$ set to zero. The results are shown in Fig. 9. During the voltage drop to 0.8 pu, in Fig. 9(a), the EAHO injects

1400 var, while Fig. 9(b) indicates that the AHO injects only 1050 var, which is 25% less than the EAHO. The results for the droop controller in Fig. 9(c) show a reactive power injection of 1450 var, slightly exceeding the EAHO's output. These results are consistent with the theoretical calculations in Section II.B, where the injected reactive powers were 1443 var, 1078 var, and 1529 var for the EAHO, AHO, and droop controllers, respectively. For the voltage jump to 1.1 pu, EAHO and droop controls absorb 600 var, while AHO absorbs 450 var. Both experimental and theoretical results confirm that the proposed EAHO provides approximately 25% enhanced grid support in terms of reactive power support compared to the AHO under both voltage sag and swell conditions.

## C. Test Scenario 3: Active power reference change

To compare the dynamic response of different controllers and validate the results from Fig. 6, $P_{ref}$ changes from 500 W to 2000 W. The results, presented in Fig. 10, confirm that all strategies successfully track the reference power. Figs. 10(a) and b) show faster transient responses for the EAHO and AHO, with a settling time ($t_{set}$) of 200 ms. However, the droop control exhibits a much slower response, with $t_{set}$=500 ms and an overshoot (OS) of 33%, as shown in Fig. 10 (c). These observations are consistent with the eigenvalue comparison in Fig. 6.

## D. Test Scenario 4: Power-sharing in stand-alone mode

In this test, both AHO-based and EAHO-based inverters are connected in parallel with a conventional droop control inverter to supply local loads in stand-alone mode. Initially, the inverters supply a resistive load $R_{L1} = 94\ \Omega$, and then an additional load $R_{L2} = 33\ \Omega$ is added in parallel. Since frequency restoration in secondary control can affect power-sharing, it is neglected in this test scenario to better demonstrate the controllers' effect on power-sharing [27]. Fig. 11 shows the experimental results, where $P_{EAHO}/i_{EAHO}$, $P_D/i_D$, and $P_{AHO}/i_{AHO}$ represent the active power and current waveforms for EAHO, conventional droop, and AHO controllers, respectively. Fig. 11(a) illustrates the waveforms when EAHO operates in parallel with the droop control. Before and after the load change, both inverters share the active power equally, each delivering 240 W and 920 W, respectively. However, for the AHO in parallel with droop control, as shown in Fig. 11(b), the active power delivered is 220 W and 260 W before the load change, and 840 W and 1000 W after the load change, respectively. This represents about a 16% power difference between the inverters, which is also reflected in the current waveforms.

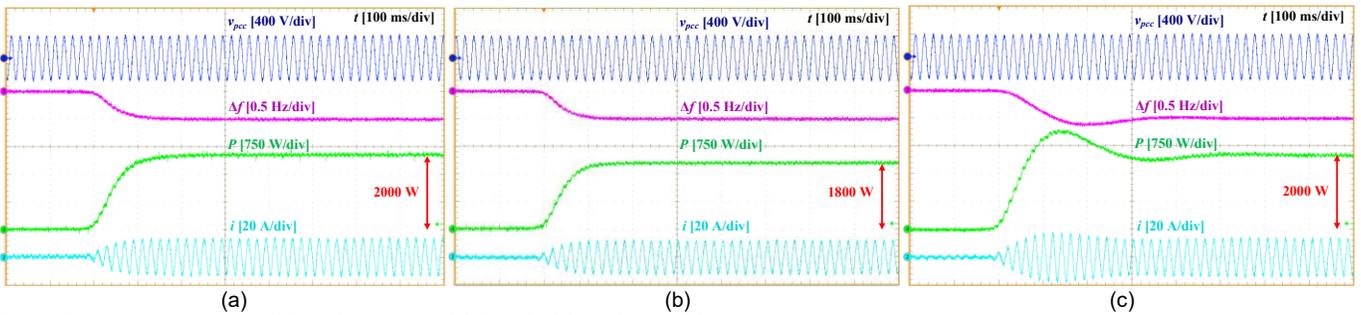

Fig. 8. Experimental results of Test Scenario 1: (a) EAHO, (b) AHO, and (c) droop controllers.

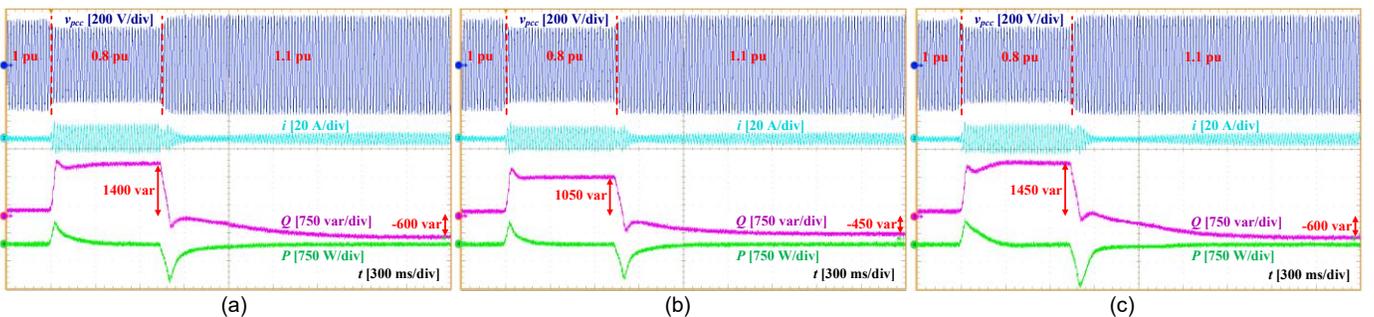

Fig. 9. Experimental results of Test Scenario 2: (a) EAHO, (b) AHO, and (c) droop controllers.

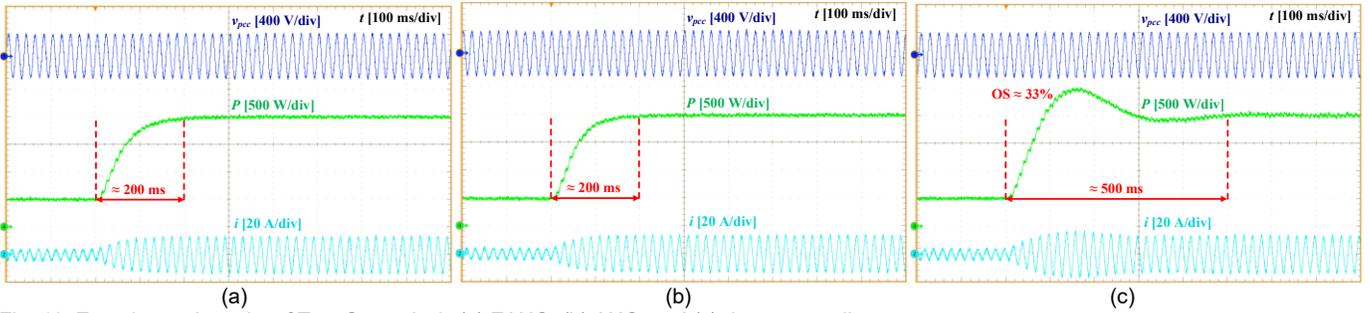

Fig. 10. Experimental results of Test Scenario 3: (a) EAHO, (b) AHO, and (c) droop controllers.

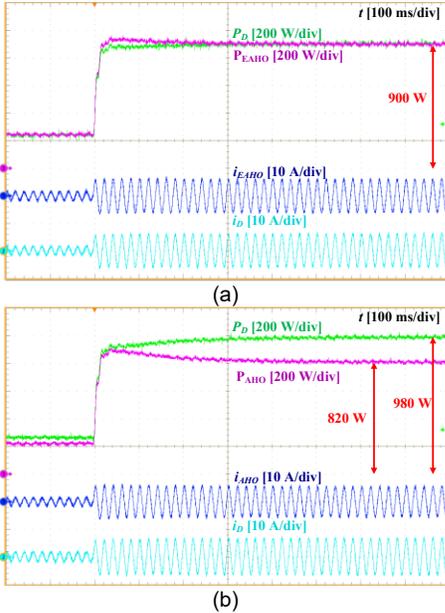

Fig. 11. Experimental results of Test Scenario 4: droop controller parallel with (a) EAHO and (b) AHO.

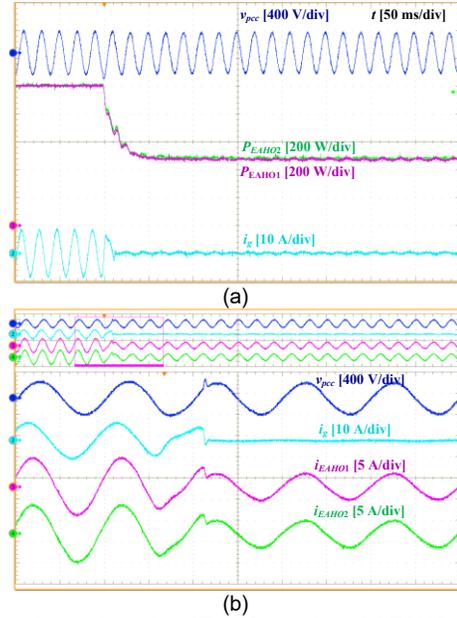

Fig. 12. Experimental results of Test Scenario 5: (a) EAHO parallel with EAHO and (b) zoomed view of current waveforms.

### E. Test Scenario 5: Grid disconnection in parallel operation

In this scenario, two inverters operating are connected in parallel to supply a local load ($R_L$=47 Ω) and the grid. Before grid disconnection, each inverter was programmed with $P_{ref}$ = 1000 W, meaning each supplied 1000 W to the combined load and grid. Figs. 12 and 13 present the experimental results when the grid is suddenly disconnected by opening its relay. In these results, $i_g$ is the grid current. Fig. 12(a) demonstrates the results when two inverters with EAHO control are connected in parallel. This figure shows that each inverter supplies 1000 W before the grid disconnection. After disconnection, the output power of both inverters shifts smoothly to 480 W to supply the local load equally. This smooth transition is evident in the zoomed-in voltage and current waveforms of Fig. 12(b). These results confirm that both inverters continued to operate without tripping, demonstrating the robustness of the proposed strategy in maintaining stable operation following grid disconnection.

To further validate the improved power-sharing accuracy of the proposed EAHO, this test scenario was repeated for the inverters with different control strategies. Fig. 13(a) illustrates the waveforms when the EAHO operates in parallel with droop control. After grid disconnection, both inverters settle at 480 W in steady state, supplying the local load with negligible power-sharing error. In contrast, Fig. 13(b) illustrates that when the AHO operates in parallel with droop control, a 16% power mismatch occurs, as expected because of the voltage-dependent APL droop coefficient of the AHO.

Table II compares the AHO, EAHO, and droop control based on experimental results. The EAHO achieves improved steady-state performance and power-sharing accuracy compared to the AHO, and successfully retains its fast dynamic response. Moreover, the EAHO exhibits a faster response and improved damping compared to droop control, while still achieving similar steady-state performance.

### F. Test Scenario 6: Variation in grid impedance and control parameters

This section examines the effect of variation in control and grid impedance parameters on the EAHO's transient performance. The results are compared with those presented for the designed values in Fig. 10(a), where $L_g$=1 mH, $R_g$=1 Ω, and $\eta_e$=$\eta_{e,set}$=0.0016. Figs. 14(a) and (b) show the EAHO's response when $P_{ref}$ jumps from 500 W to 2000 W under different grid impedance values. Fig. 14(a) demonstrates a significantly slower response (760 ms) when $L_g$ is increased to 15 mH, consistent with the eigenvalue map in Fig. 5(a). In contrast, Fig. 14(b) shows only a slight change in the transient response when $R_g$ is reduced to 0.5 Ω, which agrees with the eigenvalue map in Fig. 5(b), where the dominant eigenvalues shift only slightly with $R_g$. The controller's resilience against real-time grid impedance variations is further investigated in Fig. 14(c). In this test, the inverter operates at $P_{ref}$ = 1500 W with $L_g$ = 5 mH. Then, a 1 mH inductor is added in parallel and subsequently removed in real time, altering $L_g$ to 0.83 mH and back. The results confirm that the EAHO's steady-state tracking performance is not affected despite sudden grid impedance changes.

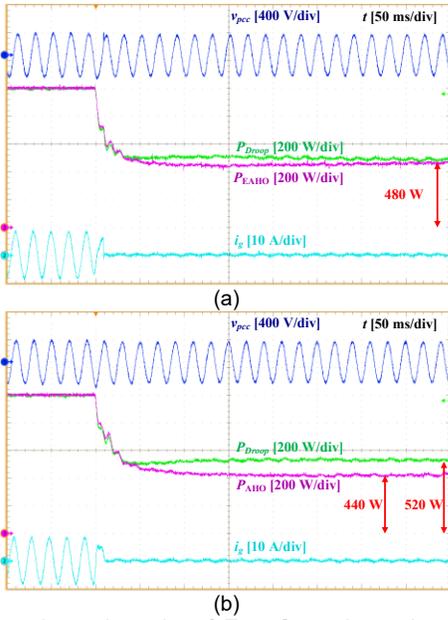

Fig. 13. Experimental results of Test Scenario 5: droop controller parallel with (a) EAHO and (b) AHO.

| GFM strategy | Active and reactive power support (Figs. 8 and 9) | Dynamic response (Fig. 10) | Power sharing accuracy (Figs. 11 and 13) |
|---|---|---|---|
| EAHO | 2000 W, 1400 var | Fast | Accurate |
| AHO [12] | 1800 W, 1050 var | Fast | 16% difference |
| Droop [1] | 2000 W, 1450 var | Slow | Accurate |

Fig. 15(a) shows the results for different control parameters, $\eta_e$=0.5$\eta_{e,set}$, when $P_{ref}$ changes from 500 W to 2000 W. Comparing this figure with Fig. 10(a) reveals that active power reaches its final value within 480 ms and 200 ms for $\eta_e$=0.5$\eta_{e,set}$ and $\eta_e$=$\eta_{e,set}$, respectively. These results are consistent with the eigenvalue map of Fig. 4(a), in which decreasing $\eta_e$ results in a slower response.

Similar to AHO, the initial synchronisation time of the EAHO is primarily influenced by $\mu_e$. Thus, to validate the small-signal analysis, the output voltage generation time is measured for different $\mu_e$ values. Figs. 15(b) and (c) present the results, showing that the oscillator's voltage reaches its final value within 280 ms and 110 ms for $\mu_e$=$\mu_{e,set}$ and $\mu_e$=4$\mu_{e,set}$, respectively. These results are consistent with the eigenvalue map in Fig. 4(b), where increasing $\mu_e$ results in a faster response.

## IV. CONCLUSION

This paper presents an EAHO control strategy for GFM inverters that effectively addresses the inherent challenges of the conventional AHO strategy. By introducing a simple yet effective modification to the current error feedback term, the proposed EAHO achieves a linear, voltage-independent active

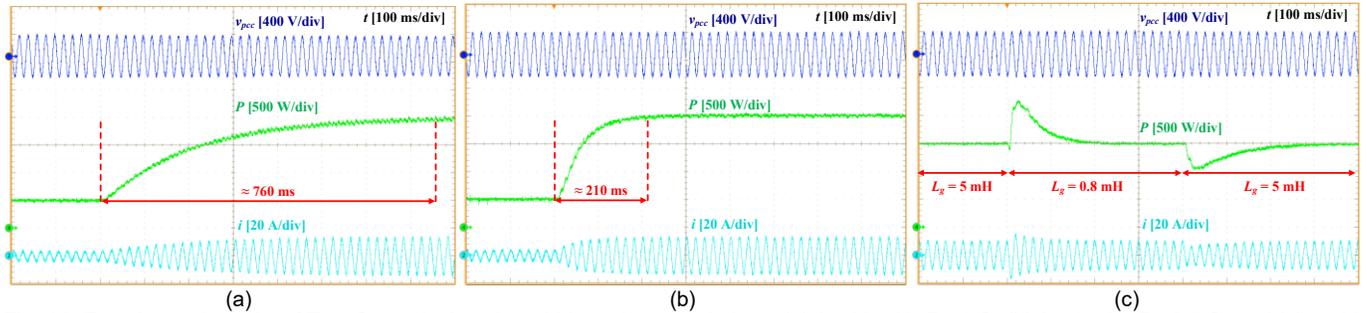

Fig. 14. Experimental results of Test Scenario 6 under grid impedance variations: (a) $L_g$=15 mH, $R_g$=1 Ω, (b) $L_g$=1 mH, $R_g$=0.5 Ω, and (c) *sudden changes in $L_g$*.

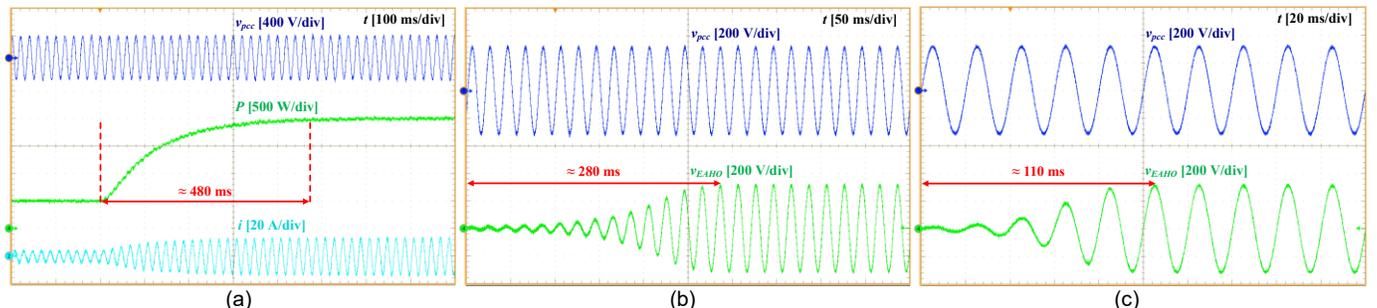

Fig. 15. Experimental results of Test Scenario 6 under different control parameters: (a) $\eta_e$=0.5$\eta_{e,set}$, $\mu_e$=$\mu_{e,set}$, (b) $\eta_e$=$\eta_{e,set}$, $\mu_e$=$\mu_{e,set}$, and (c) $\eta_e$=$\eta_{e,set}$, $\mu_e$=4$\mu_{e,set}$.

$$A = \begin{bmatrix} 2\mu_e(V_0^2 - 3V_{eq}^2) + \eta_e Q_{ref} - 2\eta_e V_{eq} F_2^* & -\eta_e V_{eq}^2 F_1^* & -\eta_e V_{eq}^2 \sin(\theta_{eq}) & \eta_e V_{eq}^2 \cos(\theta_{eq}) \\ -\eta_e F_1^* & \eta_e V_{eq} F_2^* & -\eta_e V_{eq} \cos(\theta_{eq}) & -\eta_e V_{eq} \sin(\theta_{eq}) \\ \dfrac{\cos(\theta_{eq})}{L_T} & -\dfrac{V_{eq} \sin(\theta_{eq})}{L_T} & -\dfrac{R_T}{L_T} & \omega_{eq} \\ \dfrac{\sin(\theta_{eq})}{L_T} & \dfrac{V_{eq} \cos(\theta_{eq})}{L_T} & -\omega_{eq} & -\dfrac{R_T}{L_T} \end{bmatrix} \tag{22}$$

$$A = \begin{bmatrix} -\omega_q(1 + m_q F_2^*) & -m_q \omega_q V_{eq} F_1^* & 0 & -m_q \omega_q V_{eq} \sin(\theta_{eq}) & m_q \omega_q V_{eq} \cos(\theta_{eq}) \\ 0 & 0 & 1 & 0 & 0 \\ -m_p \omega_p F_1^* & m_p \omega_p F_2^* & -\omega_p & -m_p \omega_p V_{eq} \cos(\theta_{eq}) & -m_p \omega_p V_{eq} \sin(\theta_{eq}) \\ \dfrac{\cos(\theta_{eq})}{L_T} & -\dfrac{V_{eq} \sin(\theta_{eq})}{L_T} & I_q & -\dfrac{R_T}{L_T} & \omega_{eq} \\ \dfrac{\sin(\theta_{eq})}{L_T} & \dfrac{V_{eq} \cos(\theta_{eq})}{L_T} & -I_d & -\omega_{eq} & -\dfrac{R_T}{L_T} \end{bmatrix} \tag{23}$$

power droop coefficient while preserving the core structure and dynamic advantages of the original AHO. The extensive analyses demonstrated improved active and reactive power support, enhanced droop performance, and stable operation under wide-ranging grid conditions. Experimental results validated the theoretical predictions, confirming the EAHO's improved grid support under grid frequency and voltage fluctuations and enhanced power-sharing accuracy in parallel operation, while it preserves the fast dynamic performance of the conventional AHO.

## APPENDIX A

The Jacobian matrix $A$ for the AHO and droop control are given in (22) and (23), respectively, where state variables for the AHO are $\Delta x = [\Delta v, \Delta \theta, \Delta i_d, \Delta i_q]^T$ and for the droop control are $\Delta x = [\Delta v, \Delta \theta, \Delta \omega, \Delta i_d, \Delta i_q]^T$, and $F_1^*$ and $F_2^*$ are given in (21).

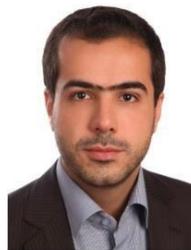

**Mohammad Monfared** (Senior Member, IEEE) received the M.Sc. and Ph.D. degrees (Hons.) in electrical engineering from Amirkabir University of Technology, Tehran, Iran, in 2006 and 2010, respectively. Since 2022, he is a Senior Lecturer with the Faculty of Science and Engineering, Swansea University, Swansea, U.K. His research interests include power electronics, renewable energy systems, and energy management. Dr. Monfared is an Associate Editor of IEEE TRANSACTIONS ON INDUSTRIAL ELECTRONICS and IEEE TRANSACTIONS ON POWER ELECTRONICS, and a member of IEEE IES Technical Committee on Renewable Energy Systems (TC-RES).

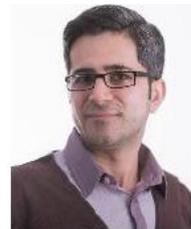

**Meghdad Fazeli** (Senior Member, IEEE) received the M.Sc. and Ph.D. degrees from the University of Nottingham, U.K., in 2006 and 2011, respectively. From 2011 to 2012, he was a Research Assistant with Swansea University, recruited on ERDF-funded project of the Solar Photovoltaic Academic Research Consortium (SPARC), where he worked on integration of large PV systems, aimed to provide ancillary services. In 2013, he worked for couple of months on Smart Operation for Low Carbon Energy Region (SOLCER) Project as a Research Officer with Swansea University. In September 2013, he became an Academic Staff with the Electronic and Electrical Engineering Department, Swansea University, where he is currently a Senior Lecturer. His main research interests include integration and control of renewable energy, ancillary services, VSMs, energy management systems, and micro/ nano-grids.

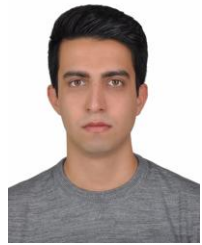

**Hamed Rezazadeh** received the B.Sc. and M.Sc. degrees (Hons.) in electrical engineering from Ferdowsi University of Mashhad, Mashhad, Iran, in 2017 and 2020, respectively. He is currently working toward his Ph.D. degree in power electronics at Swansea University, Swansea, U.K.
His research interests include power electronics, renewable energy systems, and grid-forming inverters.
Mr. Rezazadeh is an active reviewer for various IEEE TRANSACTIONS.

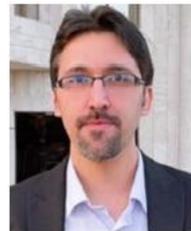

**Saeed Golestan** (Senior Member, IEEE) received the Ph.D. degree in electrical engineering from Aalborg University, Aalborg, Denmark, in 2018. He is currently an Associate Professor with AAU Energy, Aalborg University. His research interests include modeling, synchronization, and control of power electronic converters and microgrids for renewable energy systems. He is an Associate Editor for IEEE TRANSACTIONS ON POWER ELECTRONICS, IEEE TRANSACTIONS ON INDUSTRIAL ELECTRONICS, IEEE OPEN JOURNAL OF THE INDUSTRIAL ELECTRONICS SOCIETY, and IEEE JOURNAL OF EMERGING AND SELECTED TOPICS IN POWER ELECTRONICS. He was the recipient of several awards, including the 2023 IEEE TPE Associate Editor Excellence Award and the 2022 Energies Young Investigator Award.